\documentclass[12pt,preprint]{aastex}
\usepackage{amsbsy}
\usepackage{amsmath}
\newcommand{\be}{\begin{equation}}
\newcommand{\ee}{\end{equation}}

\newcommand{\vn}{{\bf v}_{n}}
\newcommand{\vns}{{\bf v}_{ns}}
\newcommand{\vs}{{\bf v}_{s}}

\newcommand{\om}{\boldsymbol{\omega}}

\newcommand{\Ro}{{\it Ro}}
\newcommand{\Ek}{{\it E}}

\def\ltsima{$\; \buildrel < \over \sim \;$}

\def\lsim{\lower.5ex\hbox{\ltsima}}
\def\gtsima{$\; \buildrel > \over \sim \;$}
\def\gsim{\lower.5ex\hbox{\gtsima}}

\begin{document}
\title{An unstable superfluid Stewartson layer in a differentially rotating neutron star}

\author{C. Peralta\altaffilmark{1,2} and A. Melatos\altaffilmark{3}}

\email{cperalta@aei.mpg.de}

\altaffiltext{1}{Max-Planck-Institut f\"ur
Gravitationsphysik, Albert-Einstein-Institut, Am M\"uhlenberg 1,
D-14476 Golm, Germany}

\altaffiltext{2}{Honorary Fellow: School of Physics, University of Melbourne.}

\altaffiltext{3}{School of Physics, University of Melbourne,
Parkville, VIC 3010, Australia}

\begin{abstract}
\noindent 
Experimental and numerical evidence is reviewed for the existence of
a Stewartson layer in spherical Couette flow
at small Ekman and Rossby numbers 
($\Ek \lsim 10^{-3}$, $\Ro \lsim 10^{-2}$), the relevant
hydrodynamic regime in the superfluid outer core of 
a neutron star. Numerical simulations of a superfluid Stewartson
layer are presented for the first time, showing how the layer
is disrupted by nonaxisymmetric instabilities. The unstable
ranges of $\Ek$ and $\Ro$ are compared with estimates
of these quantities in radio pulsars that exhibit
glitches. It is found that glitching pulsars lie on the
stable side of the instability boundary, allowing differential
rotation to build up before a glitch.
\end{abstract}


\section{Introduction}
Meridional circulation, driven by Ekman
pumping, occurs routinely in the atmospheres, oceans, and
fluid interiors of rapidly rotating astrophysical
objects. Indeed, it is a generic feature of Navier--Stokes flow
in any spherical Couette geometry (i.e., a differentially rotating,
spherical shell); see \citet{je00} for a review.
As the Ekman number $\Ek$
decreases, and the differential rotation increases, spherical
Couette flow becomes nonaxisymmetric and eventually
turbulent \citep{ntz02a}.
In the limit of rapid overall rotation,
a detached shear layer,
known as the Stewartson layer, forms
along the tangent cylinder to the inner sphere
\citep{stewartson57,stewartson66,busse68}.
It can be disrupted by nonaxisymmetric
instabilities.
Numerical simulations
indicate that a multiplicity of transition states
are thereby excited \citep{h94,dcj98,h03,sc05,hje06}.

The possible existence of an unstable Stewartson layer
in a differentially rotating neutron star has
important astrophysical consequences. This is true
especially if the inner core rotates faster
than the rest of the star, like in the Earth, a real
possibility in models where the inner core makes
a transition to a crystalline color-superconducting
phase \citep{ar03,ajkkr05,asrs08}.
Recently, the importance of the global flow pattern
inside a neutron star to the phenomenon of pulsar glitches has
been highlighted  by simulations 
on the vortex \citep{wm08} and
hydrodynamic \citep{pmgo05a} levels. 
Observational data suggest that glitches result from scale-invariant
vortex avalanches driven by differential rotation \citep{mpw08}.
If the meridional circulation is fast enough,
a vortex tangle is
alternately created and destroyed,
producing impulsive and oscillatory
torque variations \citep{pmgo06b,asc07,mp07}.
The presence of a Stewartson layer modifies
these conclusions and those of other studies (e.g., of stellar oscillations),
where a multicomponent superfluid is
perturbed starting from a nontrivial equilibrium state
\citep{pmgo06b,pmgo08,gaj09,phaj09}.

To date, no studies have been published of the formation
and stability of a Stewartson layer in {\it superfluid}
spherical Couette flow.
In this Letter, we present the first numerical
simulation of such a system as an idealized model
of the superfluid outer core of 
a neutron star. We calculate stability curves for
a range of unstable nonaxisymmetric modes
and compare the conditions for instability
with observational glitch data, finding
an upper limit
on the velocity shear and hence the
glitch sizes observed.
The paper is organized as follows. In \S\ref{sec:hydro}, we briefly review the analytic and numerical
theory of Stewartson layers in viscous fluids, before 
calculating the structure of a
steady Stewartson layer in a
$^1$S$_0$-paired neutron superfluid.
In \S\ref{sec:unstable}, we study
the stability of the layer to nonaxisymmetric perturbations
as a function of Ekman number $\Ek$ and Rossby number $\Ro$.
In \S\ref{sec:stewartson}, we compare the stability basin
in the $\Ek$-$\Ro$ plane with available glitch data.
The astrophysical implications 
are discussed in \S\ref{sec:discussion}.

\section{Stewartson layers in neutron stars}
\label{sec:hydro}
Consider a viscous fluid flowing
inside a differentially rotating spherical container,
with inner radius (angular frequency) $R_1$ ($\Omega_1$), outer radius (angular frequency) $R_2$ ($\Omega_2$), Rossby number  $\Ro = (\Omega_2-\Omega_1)/\Omega_2 \ll 1$, and
Ekman number $\Ek = \nu_n/(R_2-R_1)^2 \Omega_2$, where $\nu_n$ denotes
the kinematic viscosity. In a frame corotating with the outer sphere,
the fluid outside the cylinder tangential to
the inner sphere is at rest, while the fluid inside the tangent 
cylinder moves in a columnar fashion \citep{p56}. 
Fluid is expelled (sucked in) by Ekman
layers at $r=R_2$ ($R_1$), while a triple-deck Stewartson layer 
buffers the jump in 
angular velocity across the tangent cylinder.  
It consists of an inner layer of thickness $\Ek^{1/3}$
sandwiched between layers of
thicknesses $\Ek^{2/7}$ ($\Ek^{1/4}$) just inside (outside) the tangent cylinder \citep{stewartson66}. 
The Ekman layers scale as $\Ek^{1/2}$, except near the equator
of the inner sphere, where they scale as $\Ek^{2/5}$. 

\begin{figure*}
\epsscale{0.9}
\plottwo{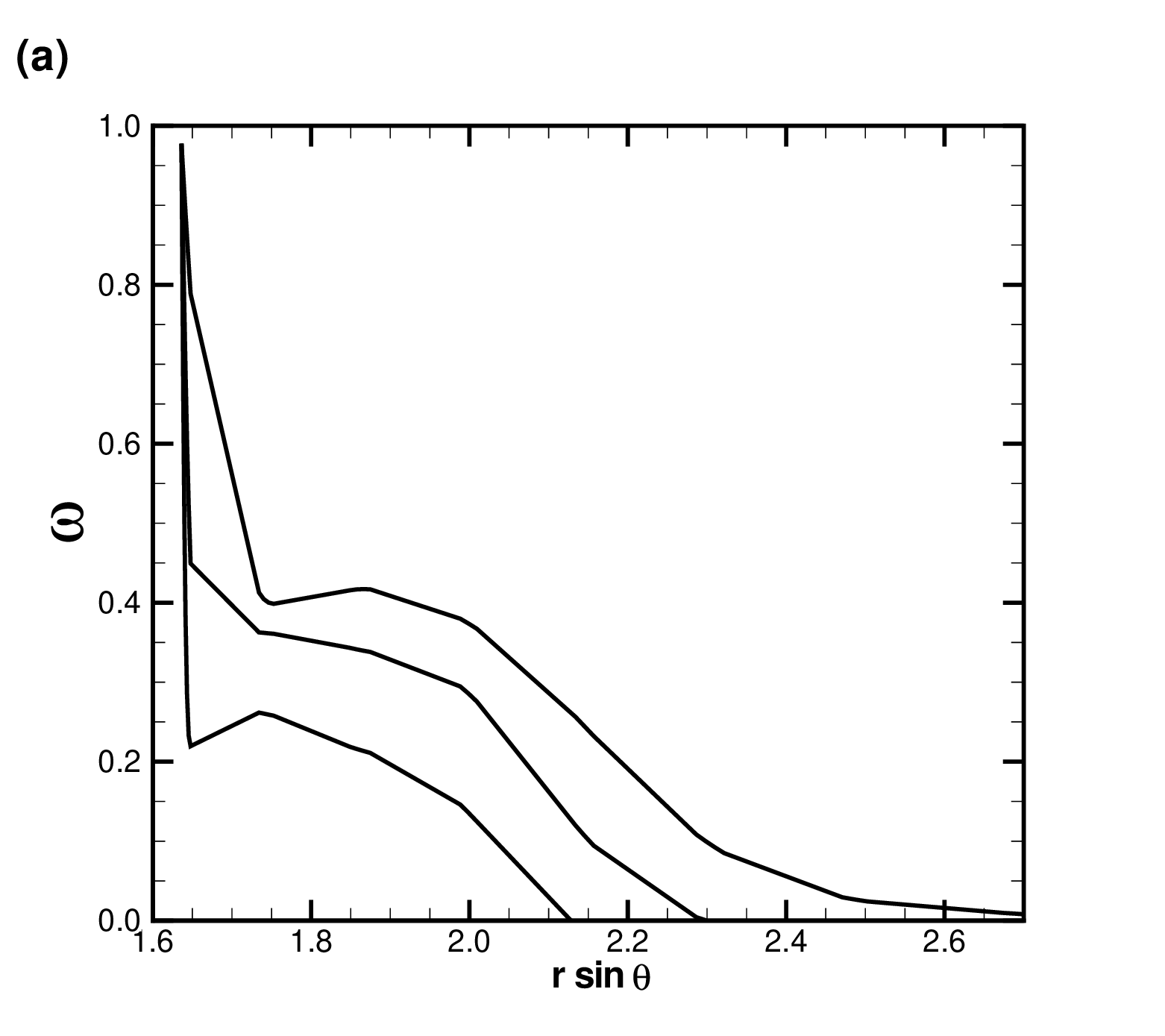}{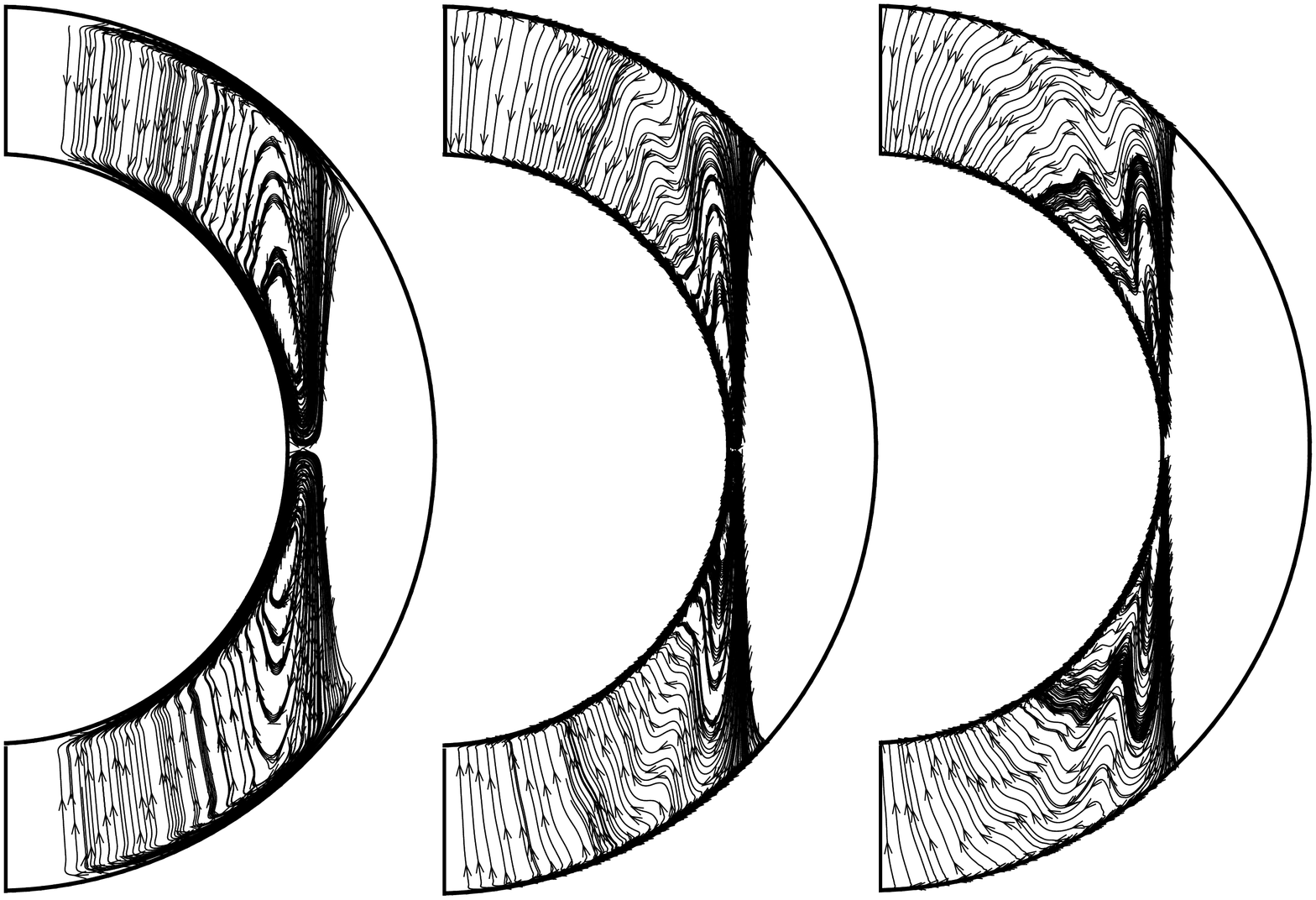}
\caption{(a) Angular velocity $\omega$ as a function of cylindrical
radius $r \sin \theta$, at $z = 1.2$, for the normal fluid
component in superfluid spherical Couette flow, for
$\Ek = 1 \times 10^{-3}$ (upper curve), $1 \times 10^{-4}$ (middle curve), and $2 \times 10^{-5}$ (lower curve). (b) Meridional streamlines of the normal component for 
$\Ek = 1 \times 10^{-3}$ (left), $1 \times 10^{-4}$ (center), and $2 \times 10^{-5}$ (right). $\Ro$ equals $10^{-4}$ in all plots.}
\label{fig:stew1}
\end{figure*}

A superfluid Stewartson layer in a spherical
shell exhibits a similar structure.
Figure \ref{fig:stew1}a graphs the angular velocity in
the rotating frame as a function of the cylindrical radius $s = r \sin \theta$,
for $\Ek = 1 \times 10^{-3}$ (upper curve), $ 1 \times 10^{-4}$ (middle curve),
and $2 \times 10^{-5}$ (lower curve), with $\Ro=10^{-4}$. The layer starts
at cylindrical radius $s \approx 1.8$ 
and its thickness decreases with decreasing $\Ek$,
extending out to $s \approx 2.7$ for $\Ek=1 \times 10^{-3}$ 
and $s \approx 2.1$ for $\Ek=2 \times 10^{-5}$.
In viscous flows,
the thickness of the layer changes by less than $1$ \%
for $0 \lsim \Ro \lsim 0.5$ \citep{h03}; similar
behaviour is observed here.
The inner Ekman layers are thicker at the equator.

Figure \ref{fig:stew1}b displays meridional streamlines
for $\Ek = 1 \times 10^{-3}$ (left), $ 1 \times 10^{-4}$ (center)
and $2 \times 10^{-5}$ (right), with $\Ro= 10^{-4}$.
The Stewartson layer is visible along the tangent cylinder, narrowing
from left to right.
The streamlines are drawn in the rotating frame
of the outer sphere; the blank region to the right of
the tangent cylinder indicates that the fluid there is at rest.
The characteristic meridional speed in the layer scales as $E^{0.1} (R_2 - R_1) \Omega_2$.

To obtain the results in Figure \ref{fig:stew1}, we solve the two-component Hall-Vinen-Bekarevich-Khalatnikov
(HVBK) equations for a superfluid inside
a spherical differentially rotating shell, with $R_1/R_2 = 0.67$,
using a pseudospectral
collocation and time-split method \citep{pmgo05a,pmgo08}.
The details of the
calculation will be set out in a longer paper.
Boundary conditions assume the presence of an
inner core or a transition between
a $^1S_0$ and $^3P_2$ superfluid \citep{yls99,mm05}. 
We ignore vortex pinning and proton-neutron entrainment 
for simplicity, although recent work shows it to
be important \citep{ss95,ac01}.
We adopt no-slip and no-penetration boundary conditions for the normal
fluid component (velocity $\vn$) and no-penetration
for the superfluid component (velocity $\vs$), ignoring the small
tension force to reduce 
the order of the equation for $\vs$ by one [see \citet{hbj95}
and \citet{pmgo08} for a discussion]. The mutual friction force is 
taken to have the
anisotropic Hall-Vinen form 
[$\propto \hat{\om}_s \times \om_s \times \vns$,
with $\vns = \vn - \vs$ and $\om = \nabla \times \vs$
\citep{hv56a,hv56b}], with $B=10^{-2}$, and $B^\prime=10^{-4}$
\citep{asc07}. We take $\rho_n/\rho = 0.01$ and $\rho_s/\rho = 0.99$
for the normal and superfluid mass density fractions respectively, where
$\rho = \rho_n + \rho_s$ denotes the total density \citep{pmgo05a}.

\section{Nonaxisymmetric instabilities}
\label{sec:unstable}
The Stewartson layer is disrupted when $\Ro$ exceeds a threshold
$\Ro_c$($\Ek$), which decreases as $\Ek$ increases, exciting
a Kelvin--Helmholtz--type instability.
\citet{h03} and \cite{sc05} investigated thoroughly
the most unstable modes of a viscous
fluid Stewartson layer.
They discovered nonaxisymmetric instabilities for azimuthal
modes $ 1 \leq m \leq 119 $, with $0 \leq \Ro_c \leq 0.6$ and $10^{-10} \leq \Ek \leq 10^{-3.5}$. The azimuthal mode number of the most unstable mode
increases with decreasing $\Ek$. \citet{h03} discovered
empirically the scaling $\Ro_c \sim 0.6 \Ek^{0.65}$ for $\Ro >0$.
For $\Ro < 0$, the most unstable mode is almost
always $m=1$, with $m=2$ being excited
in the range $10^{-0.25} \lsim \Ro_c \lsim 0.1$.
In this regime, \citet{h03} found $\Ro_c \sim 0.8 \Ek^{0.45}$ .
The asymmetry with the sign of $\Ro$ is still not understood physically \citep{h03}.

Here, we generalize the numerical calculation of
$\Ro_c$ to a superfluid Stewartson layer. We follow
a two-stage recipe. First, 
for fixed
$\Ro = 10^{-4}$, we calculate axisymmetric HVBK basic
states for $\Ek = 1 \times 10^{-3}$, $ 1 \times 10^{-4}$, $5 \times 10^{-5}$, $2 \times 10^{-5}$,
and $1 \times 10^{-6}$. Second, we linearize the HVBK equations around the base state
and test the stability of a given $m \neq 0$ perturbation (typical amplitude $\approx 1$ \%) by
increasing $\Ro$ until the mode grows exponentially.
We obtain the scaling
\begin{equation}
\Ro_c \approx 4.1 \Ek^{0.40}
\label{eq:sscale}
\end{equation}
for $10^{-6} \leq \Ek \leq 10^{-3}$ and
$0.02 \leq \Ro_c \leq 0.33$. 
We concentrate our efforts on
$m=6$, the most
unstable mode at $\Ek \approx 10^{-3}$ for
viscous fluids (and also for the superfluid). 
It is important to note that $m=6$ is not the most unstable
mode at smaller Ekman numbers 
(e.g. $\Ek = 1 \times 10^{-6}$, where $m=10$ is more unstable), but
the critical Rossby number is unaffected ($\Ro_c = 0.02$).
In \S\ref{sec:unstable}, we extrapolate
the scaling (\ref{eq:sscale}) to the regime
$\Ek \gsim 10^{-12}$, $\Ro \lsim 10^{-4}$, relevant to radio
pulsars \citep{lss00,mp07}, where
numerical simulations are too hard to do.
A more thorough parameter survey
will be presented in a future paper. 

\begin{figure*}
\epsscale{0.7}
\plotone{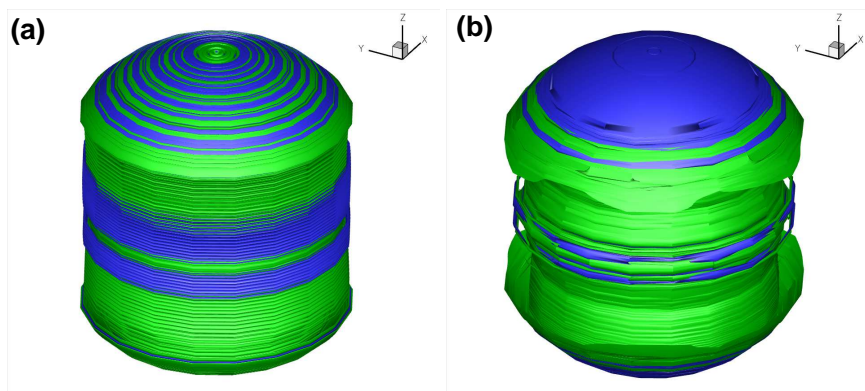}
\caption{Isosurfaces of the velocity gradient discriminant (see text): $D_A = 10^{-3}\, \Omega_2^{6}$ (blue) and $-10^{-3}\, \Omega_2^{6}$ (green) for  
$E=10^{-3.1}$ and the transition
$Ro=0.2$ $\rightarrow$ $0.3$, where $m=6$ is the most unstable mode. 
(a) $t=0$ (before transition); (b) $t=400$ (after transition).}
\label{fig:topol}
\end{figure*}

Nonaxisymmetric instabilities
can decrease the shear inside a viscous Stewartson layer by 30 \% \citep{hfme04}. We observe 
a similar but
less pronounced effect in the superfluid, e.g. when the mode $m=6$ is excited at $\Ro_c=0.3$, with $\Ek = 10^{-3.1}$. Figure \ref{fig:topol} describes the topology 
of the flow in the layer,
before (Figure \ref{fig:topol}a, $\Ro =0.2$) and after 
(Figure \ref{fig:topol}b, $\Ro=0.3$) the mode $m=6$
is excited. 
The discriminant $D_A = Q_A^3 + 27 R_A^2 /4$, with $R_A= {\rm det}(A_{ij})$, 
$Q_A = (A_{ij}^2 - A_{ij} A_{ji})/2$, and
$A_{ij}=\partial v_i /\partial x_j$,
distinguishes between regions that are focal ($D_A > 0$, blue)
and strain-dominated ($D_A < 0$, green) \citep{cpc90}.
The maximum shear changes from $d\omega /ds = 1.1$ at $t=0$
to $d\omega /ds = 1.0$ at $t=400$ (cf. viscous fluid,
where the observed change is from $d\omega /ds = 1.3$ to $0.8$).
The hexagonal flow structure also boosts the torque
one must exert on the inner sphere to maintain the shear,
although the increase is less dramatic ($\sim 10$ \%).

\section{Pulsar glitches}
\label{sec:stewartson}
As $\Ek$ and $\Ro$ control 
the stability of the Stewartson layer,
it is interesting to test
whether the amplitude and rate of incidence
of rotational irregularities like glitches are
related to these two dimensionless quantities. 
In order to calculate $\Ek$, we need to know how 
$\nu_n$ depends on the density $\rho$ and temperature $T$
in the outer core. Using the neutron-neutron scattering
viscosity formula derived by \citet{cl87}, we find
\begin{equation}
\Ek = 2.6 \times 10^{-12} 
\left( \frac{T}{10^8 \, {\rm K} } \right)^{-2}
\left( \frac{\Omega_2}{10^2 \, {\rm rad} \, {\rm s}^{-1} } \right)^{-1}
\end{equation}
with $\rho = 2.8 \times 10^{12}$ {\rm g} {\rm cm}$^{-3}$.
The core temperature $T$ is related to the surface
temperature $T_s$, e.g., via the two-zone heat-blanket
model of \citet{gpe82}, which gives 
$T/10^8 \, {\rm K} = 1.29 ({T_s}/{10^6 \, {\rm K}})^{1.8}$.
We estimate $T_s$ from
the characteristic age $\tau_c = \Omega_2/2 |\dot{\Omega}_2|$, 
combined with theoretical
cooling curves for $\tau_c \leq 10^6$ {\rm yr} \citep{page98} and
standard neutrino cooling; similar $\Ek$ distributions
are obtained with
nonstandard cooling \citep{mp07}.

In Figure \ref{fig:RoEglitch}, we plot as points the maximum $\Ro$ 
in $55$ glitching pulsars with $\tau_c \leq 10^6$ {\rm yr},
taken from Table 1 in \cite{mpw08}, assuming conservatively that
$\Ro$ is less than the 
fractional frequency jump of the largest glitch.
For each object, $\Ek$ is estimated by the method in the previous paragraph. 
We also plot two $\Ro_c$ ($\Ek$) curves: the HVBK scaling computed in \S\ref{sec:hydro}
($\Ro_c \approx 4.1 \Ek^{0.40}$, solid curve) and
a scaling extrapolated from the study by \citet{sc05}
for $\Ro_c \lsim 10^{-2}$ and $\Ek \lsim 10^{-5}$
($\Ro_c \approx 9.4 \Ek^{0.57}$, dashed curve).
The viscous fluid scaling 
lies in the middle of the cluster of points,
while the HVBK superfluid scaling lies above all the points.
This suggests two possible conclusions: (i) a glitching pulsar must have
$\Ro < \Ro_c (\Ek)$, otherwise Stewartson layer instabilities would
erase the shear required for the glitch phenomenon to occur; and (ii) it is
important to include the HVBK superfluid dynamics to ensure that
all points in Figure \ref{fig:RoEglitch} lie below the $\Ro_c (\Ek)$ curve,
given that no discernible difference is observed in the glitch behaviour
of objects above and below the viscous fluid scaling.

In the superfluid Stewartson layer, there is an extra ingredient
that influences the instability curve:
the mutual friction between the normal and superfluid component, controlled by the dimensionless parameters
$B$ and $B^\prime$. We have only considered the weak coupling
regime in this investigation ($B=10^{-2}$, $B^\prime=10^{-4}$).
In the strong coupling regime [$B=0$, $B^\prime=1$ \citep{asc07}]
the instabilities are likely to be quite different. Preliminary
results for axisymmetric steady states show
that the Stewartson layer is $\sim 10$ \% thicker.
A more detailed investigation of the effect of mutual friction
and entrainment on $\Ro_c (\Ek)$ will be presented in a future paper.

\section{Discussion}
\label{sec:discussion}

The results in \S\ref{sec:hydro} and \S\ref{sec:unstable} demonstrate 
that Stewartson layers develop in rotating HVBK superfluids, like the
$^1S_0$-paired neutron superfluid in the outer core
of a neutron star. We present the first numerical simulation of
a HVBK Stewartson layer, for Ekman and Rossby numbers in the ranges
$10^{-6} \leq \Ek \leq 10^{-3}$ and $10^{-4} \leq \Ro \leq 10^{-2}$.
Superfluid Stewartson layers
are unstable to nonaxisymmetric perturbations. 
Transitions between unstable Stewartson
modes in the superfluid 
are different to those found in a viscous fluid, and the
critical Rossby number $\Ro_c (\Ek)$ is $\sim 10$ times higher in
the above parameter regime. 
In glitching pulsars, one finds $\Ro < \Ro_c (\Ek)$ in the 55
pulsars for which $\Ro$ and $\Ek$ can be estimated reliably
from observations. One possible interpretation of these data is that the Stewartson layer remains stable in these objects,
allowing rotational shear to build up (as required for glitches) without
triggering disruption of the Stewartson layer [which would
nullify the shear  \citep{hfme04} and hence shut down
the glitching behaviour].
The threshold $\Ro_c (\Ek)$ can be compared against the upper limit
derived independently from the gravitational-wave
spin down caused by Kolmogorov-like superfluid turbulence
excited in the stellar interior \citep{mp09}.

\begin{figure*}
\epsscale{0.7}
\plotone{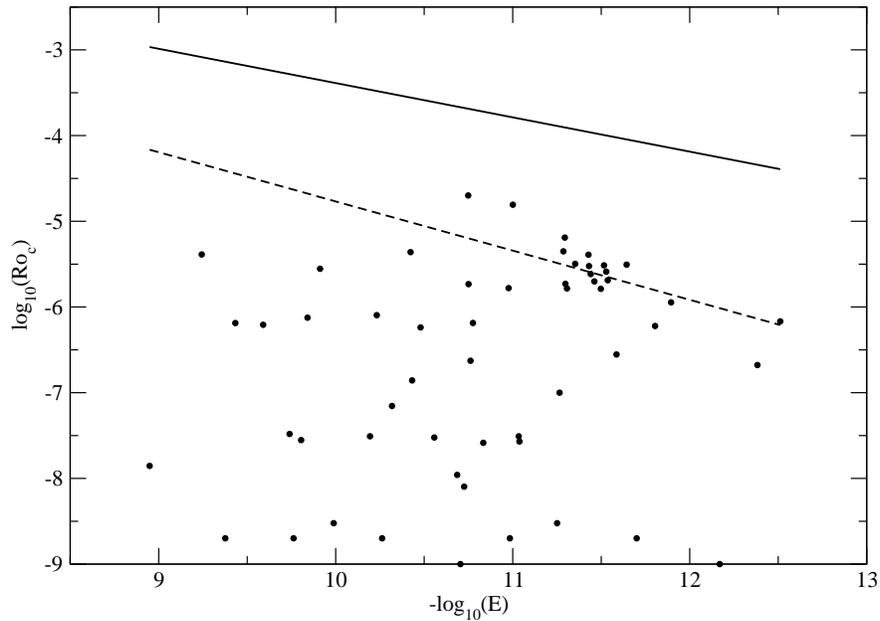}
\caption{Critical Rossby number $\Ro_c$ versus Ekman number $\Ek$ for 55 glitchers
with characteristic ages $\tau_c \leq 10^6$ yr. The points indicate
upper limits on $\Ro_c$ derived from the largest observed glitch in each object.
$\Ek$ is calculated using standard neutrino cooling. 
The solid curve is a fit to the HVBK superfluid simulations in \S\ref{sec:hydro} and \S\ref{sec:unstable}.
The dashed curve 
is an extrapolated fit to the viscous fluid calculations from \citet{sc05}.}
\label{fig:RoEglitch}
\end{figure*}

The conclusions drawn from the data in Figure \ref{fig:RoEglitch}
extend and partially clarify the surprising empirical finding,
that the $\Ek$ distribution 
is markedly different between glitchers and nonglitchers \citep{mp07}.
It seems strange that stars with
$\Ek \sim 10^{-10}$ (few glitches) 
and $\Ek \sim 10^{-12}$ (many glitches) behave
so differently, since
in both regimes Kolmogorov turbulence must be
fully developed and scale-free. However, from Figure \ref{fig:RoEglitch}
[and Figure 4 in \citet{h03}]
one sees that small differences in $\Ek$ and $\Ro$
lead to very different flow states and
stability properties \citep{h03,hfme04}.
Moreover, if theoretical estimates of effective viscosity
and hence $\Ek$ are too low by
a factor $\sim 10^5$, due to turbulent Reynolds stresses
\citep{mp07}, glitching pulsars lie in a range where
(i) the most interesting flow
transitions occur before the flow becomes turbulent
(i.e. at $10^{-4} \lsim \Ek \lsim 10^{-2}$ for $1 \leq m \leq 6$ ), and
(ii) numerical simulations 
are computationally tractable.

The results of this paper
{\it do not} prove that Stewartson flow transitions control glitch behaviour.
We merely find empirically that all observed glitchers lie on the 
stable side of the $\Ro_c (\Ek)$ threshold for
nonaxisymmetric instabilities of a Stewartson layer in a HVBK superfluid.
Recently, a hydrodynamic
trigger for pulsar glitches was proposed by \citet{ga09},
associated with r-modes excited by differential rotation.
It would be interesting to see how meridionally circulating Stewartson base
states modify these calculations, especially in the strong pinning
scenario (for which $B^\prime = 1$, $B=0$).
\citet{pls08} found empirical
evidence of departures from solid-body rotation in a radio pulsar.

We do not consider stratification or magnetic fields in this investigation.
In viscous fluids, strong stratification supresses the Stewartson layer,
leaving a Taylor column parallel to the rotation axis \citep{h09}.
Magnetic fields widen the Stewartson
layer, merging it with the interior flow \citep{h94}.
This difficult physics deserves investigation.

\acknowledgments 
We acknowledge the
support of the Max-Planck Society
(Albert-Einstein Institut) and
the computer time supplied by the Victorian Partnership for
Advanced Computation. 
We thank Rainer Hollerbach
for indicating the importance of the Stewartson layer
and very illuminating discussions.
This work made use of the ATNF Pulsar Catalogue
[http://www.atnf.csiro.au/research/pulsar/psrcat; 
\citet{atnf_catalog_05}].


\end{document}